\documentclass{singlecol-new}

\usepackage[round]{natbib}
\usepackage{stfloats}
\usepackage{mathrsfs}
\usepackage{graphicx}
\usepackage{hyperref}
\usepackage[figuresright]{rotating}
\usepackage{url}
\usepackage{color}

\theoremstyle{TH}{

}

\theoremstyle{THrm}{

}

\theoremstyle{THhit}{

}

\makeatletter
\def\theequation{\arabic{equation}}

\makeatother

\begin{document}%
\setcounter{page}{1}

\LRH{Y. Wang}

\RRH{A novel soft keyboard for touchscreen phones: QWERT}

\VOL{x}

\ISSUE{x}

\PUBYEAR{2013}

\BottomCatch


\PUBYEAR{2013}

\subtitle{}

\title{A novel soft keyboard for touchscreen phones: QWERT}

\authorA{Yanshan Wang}
\affA{School of Industrial Management Engineering,\\ Korea University,\\ Seoul, 136-713, Republic of Korea \\
E-mail: yansh.wang@gmail.com}

\begin{abstract}
The popularity of touchscreen phones has been growing around the world since the iPhones and Android phones were released. More and more mobile phones with large touchscreen have been produced, however, the phones with small size displays are still in the majority of touch phones. The foremost interface on touch smartphones is the information input module using soft keyboards. Traditional input methods on touch phones have either too small key buttons (such as QWERTY) or too many functions (such as 3$\times$4 keyboard), which are inconvenient to use. Moreover, the conventional soft keyboards only use tapping to input texts while current touch smartphones allow various gestures on the touchscreen, such as sliding. In this paper, a novel soft keyboard called QWERT is proposed for touchscreen-based smartphones. The users can interact with phones via finger gestures of tapping or sliding when input text by using the QWERT. In doing so, the interactions between users and smartphones will be faster and easier. An experiment carried out on inexperienced human subjects shows that they can learn very fast due to their familiarities with QWERTY. A simulation experiment based on a cognitive architecture, ACT-R, was also conducted to predict the movement time (MT) of experienced human subjects. The simulation results show that the MT using QWERT outperforms other default keyboards. These outcomes imply that the novel QWERT is a viable option for touch smartphone users. Based on the novel design, an application is released on Android systems. This application is expected to give better user experience for customers who use touch smartphones.
\end{abstract}

\KEYWORD{QWERT; touchscreen; smartphone; soft keyboard; Android.}

\REF{to this paper should be made as follows: xxxx (xxxx) `xxxx',
{\it xxxx}, Vol.~x, No.~x, pp.xxx--xxx.}

\begin{bio}
Yanshan Wang received the BE degree in computer science from Harbin Institute of Technology and the ME degree in management engineering from Korea University, in 2010 and 2012, respectively. He is currently working toward the PhD degree (CSC Scholarship) in the Department of Management Engineering, Korea University. His mean research interests include Nonlinear Programming, Information Retrieval and Machine Learning.
\end{bio}

\maketitle

\section{Introduction}

Since 1982 when \citet{shneiderman1982future} first used the term \textit{direct manipulation} which referred to a highly usable system using manual actions rather than typed instructions in graphical interfaces, much progress has been made in extending the input devices and techniques for direct manipulation. The development of \textit{touchscreens} is both evolutionary and revolutionary for the future of direct manipulation in Human-Computer Interaction (HCI) research area. The interaction with touchscreens, on which information display and control are one surface, is literally the most direct style of HCI \citep{albinsson2003high}. \\

\begin{figure}
  \centering
  \caption{Penetration of touchscreen technology into cellphones from 2005 to 2012 \citep{stranal}.}\label{fig.intro}
  \includegraphics[width=\textwidth]{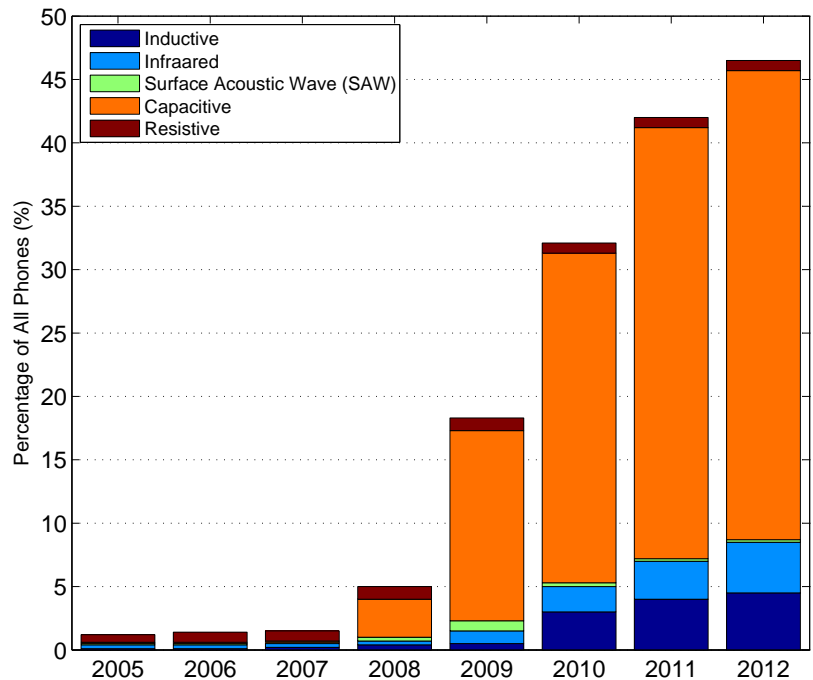}
\end{figure}

Nowadays, the touchscreens are operated everywhere, in public information kiosks, city guides, or display boards in bus stations. The increase of communication and amusement equipment, notably smartphones such as Android phones and iPhones, gives birth to the touch interfaces. According to a recent report by \citet{nielsen}, more than three out of five (61\%) mobile subscribers in the U.S. owned a smartphone during the most recent three-month period (March--May 2013). Among the smartphone owners, about 53\% used the Android OS, and 40\% the iPhones till the end of May 2013. It is interesting that nearly all the smartphone operating systems (OS) are developed based on the touchscreen. The increasing use of smartphones has given rise to the increasing use of the touchscreens. A prediction by \citet{stranal} said that the touchscreen user interfaces in mobile phones would start to see significant growth by the end of 2007 and would be used in 40\% of all phones by 2012. However, the reality has gone farther than the prediction according to the statistics shown in Figure \ref{fig.intro}. \\

One of the main services provided by smartphones is Message Service (MS), the most frequently used application. With the development of wireless network, instant messaging applications for smartphones have drawn interests recently, such as LINE, WeChat, and KakaoTalk. Therefore, the information input modules, soft keyboards, become the foremost interface on touch phones. Traditional soft keyboards are designed by simulating physical keyboards which people are familiar with. For example, most smartphone OSs embed the QWERTY as default input method. However, the layout of traditional soft keyboards, like QWERTY, is inaccurate and unsatisfying to use because the performance drawbacks of touchscreen interfaces are exaggerated with small buttons \citep{parikh2012negative}. Moreover, such soft keyboards do not make full use of the properties of touchscreens. In other words, the touchscreens share the benefits of human gestures, such as tapping and sliding, while the traditional physical keyboards are triggered merely by physical tapping. On the other hand, the smartphone users tend to use only one hand when they operate on the touchscreen. That is, they hold a smartphone with one hand and operate with a thumb. They use both hands only when the software makes one hand interaction impossible or difficult \citep{park2010touch}. Thus, the physical keyboards, such as QWERTY designed for two-hand tapping, do not fit the touchscreen devices properly. \\

Researchers have been seeking out new layouts of soft keyboards to optimize user experience on touch phones. Among those layouts, 3$\times$4 keyboards and free organized keyboard are two successful designs. However, these approaches ignored the natural characteristics of touchscreen-based smartphones. This problem motivates us to design new soft keyboards for touchscreen-based smartphone users. In our consideration, the soft keyboard design is based on the following facts:
\begin{enumerate}
  \item Most touchscreen-based smartphones are assembled with small-sized touchscreens (less than 5 inches).
  \item Users are accustomed to using thumbs to touch on smartphones.
  \item The gesture of sliding is straightforward on touchscreen-based smartphones.
\end{enumerate}

In this paper, a novel soft keyboard called QWERT is proposed for touchscreen-based smartphones. In this novel design, we take advantage of people's familiarity with traditional physical keyboards so that users are accustomed to the QWERT quickly. We also make use of the tactile characteristics of touchscreens to design the layout in order to enable users to use the gestures of tapping and sliding. Furthermore, the size of each key button is rigorously arranged. An implementation of the QWERT is developed based on an Android operating system. By using this implementation, the experiment on inexperienced human subjects is conducted in order to verify whether users can learn the QWERT fast. The experiment on experienced users is impracticable since the QWERT is a new device design. Thus, a cognitive architecture called ACT-R is used as a simulated experiment.  \\

This paper is organized as follows. The background and related research of soft keyboard design is presented in Section 2. In Section 3 of this paper, a detailed description of the design is proposed. Two experiments as well as results are established in Section 4 for inexperienced and experienced users. The last section concludes this paper with some discussions.

\section{Background and Related Research}

Among direct interaction with touch phones, bare-handed pointing is undoubtedly an intuitive way for users \citep{albinsson2003high}. The operations by bare hands are more robust and convenient, especially in special environments like outer space. As an important interface with touch devices, information input method has drawn researchers' interest since the 1980s when researchers in accessibility and Human-Computer Interface have realized the shortcoming of QWERTY as a soft keyboard layout \citep{colle2004standing, park2010touch}. They designed a variety of optimized soft keyboard layouts with increasingly more sophisticated and more rigorous optimization methods, however, the soft keyboard design is still a big challenge for touchscreen-based smartphones. \\

The starting point of soft keyboard designs is simulating the de facto physical keyboard standard layout---QWERTY, which is being operated at near maximal speeds \citep{noyes1983qwerty}. Unfortunately, QWERTY performs poorly as a soft keyboard due to the large number of buttons, which results in that each button is extremely small on the touchscreen. For example, LG LU6800, a smart phone with a 4.3 inch (94 mm$\times$54 mm) display, employs the conventional QWERTY keyboard. The size of each button is roughly 5.1$\times$7.9 mm. As a result, many studies have been focused on designing soft keyboards with the optimal sized buttons. \\

The 3$\times$4 keyboard has been the de facto standard input device for cellphones in the 1990s when cellphones became popular. Later it was transplanted to touchscreen-based smartphones. Since the alphabetical 3$\times$4 layout is not optimal for touch phones, many improvised layouts of 3$\times$4 keyboard have been proposed, such as \citet{al2011improving}, \citet{hwang2005qwerty} and \citet{mittal2009improvised}. \citet{al2011improving} proposed a new mobile text entry environment for the Arabic language based on the multi-tap text entry method. Though the improvement is obvious in the text entry method, the models are specific for the Arabic text environment. \citet{mittal2009improvised} repositioned alphabets on the 3$\times$4 keyboard to reduce the number of matches for any possible numeric combinations. Their model groups commonly used key combinations together and requires less tapping for commonly used alphabets. However, the repositioning of the board may be less familiar to users, which requires long learning time. Furthermore, the 3$\times$4 layout cannot play the role of a "full speed" keyboard due to the multi-character layout (\citep{hwang2005qwerty}). \\

Many QWERTY-like models have been developed since researchers realized that keyboard design should take people's custom and familiarity into account. \citet{bi2010quasi} showed that a quasi-QWERTY optimized soft keyboard could significantly reduce novice users' visual search time due to users' familiarity with QWERTY . \citet{matias1994half}proposed a one-handed keyboard called "Half-QWERTY", which used only half of the QWERTY keyboard and the other half was mapped onto this half. When a key was depressed, the character in the upper left of the key was entered. When preceded by holding down the space bar, the character in the lower right was entered. However, the design has to consider different models for left handed and right handed people. In addition, the "mirror" based layout is no longer applicable since most people use only left or right thumbs instead of whole fingers. As shown in these designs, the fact that people are familiar with QWERTY should be taken into consideration for new designs. \\

Another important issue for soft keyboard design is the size of each button. This issue is well-studied on physical keyboards in previous researches. In general, it is difficult to point at targets that are smaller than the finger width \citep{albinsson2003high}. The button size, larger than finger width, is preferred, however, larger size leads to larger device size. Thus many approaches tried to find the optimal minimum size of each button \citep{beringer1990target, hall1988factors}. \citet{pfauth1981person}, the pioneers to identify key size, recommend that keys should have a minimum size of 22.1 mm. \citet{beaton1985effects} implemented menu entry with the requirement of a single touch. A 4x3 matrix of keys with the size of 10x20 mm performed the best. \citet{martin1988configuring}found that a 13x13 mm key size performed better than 6x13 or 13x6 mm key sizes when users input one to three digits on a keyboard. \citet{sears1991improving}estimated an envelope of 26 mm for seated entry using a 99\% capture criterion. \citet{bender1999touch} found better performance for 30mm and 10mm square buttons when users were standing to input digits in a numeric keyboard. Recent research by Parhi et al. showed that the minimum size should range from 9.2 to 9.6mm for one-hand thumb \citep{parhi2006target}. However, \citet{darroch2005effect} reported that users could read text easily on a target of about 2 to 3 mm .\\

A recent study by \citet{colle2004standing} recommend that, key size no smaller than 20 mm is always sufficient for optimal performance or for user satisfaction on a touch entry. In addition to that recommendation, 1 mm edge-to-edge spacing should be used if sufficient space is available. We note that this study above is carried out for interaction with an index finger on PDA. The size of PDAs is much larger than touch phones. For example, the screen of LG LU6800 is 4.3 inch (93.98 mm$\times$54.102 mm) display. Obviously, the recommendation above cannot be applied in our design. A recent study concludes that the touch key size of 10 mm for mobile phones produces the best results in terms of task completion time, number of errors, and subjective satisfaction if all the measures are considered together \citep{park2010touch}. Thus this recommendation is acceptable for our design.\\

As for examining the performance of a HCI, integrating theory, data and knowledge about cognitive psychology and human performance is important guidance. The cognitive architectures is "a scientific hypothesis about those aspects of human cognition that are relatively constant over time and relatively independent of task." \citep{gray1997introduction}. \citet{card1983psychology} used an empirical model called Keystroke-Level Model (KLM) which consists of a set of actions: $MT=T_k+T_p+T_h+T_d+T_m+T_r$, where $k$, $p$, $h$, $d$, $m$, and $r$ represent the keyboard, pointing, homing, drawing, mental preparation and system response, respectively. Fitts' Law \citep{fitts1954information} predicts the time taken to move to an item using a pointing device or finger: $MT=a+b\times ID$, where $MT$ is the movement time, $a$, $b$ are constants empirically determined through regression analysis. The $ID$, representing the "index of difficulty" of the task, is defined as $ID=\log_2(A/W+1)$, where $A$ is the distance between two targets and $W$ is the width of the targets. However, these methods are low-level psychomotor models which are unable to simulate recent designs that contain complicated gestures. A number of higher-level models is frequently used in recent studies, such as LICAI/CoLiDeS, EPIC, and ACT-R. LICAI/CoLiDeS \citep{kitajima1997comprehension} is a primary example of an HCI-oriented cognitive architecture not based on the production system framework. EPIC \citep{kieras1997overview} has been used to model more complex tasks, such as menu selection. ACT-R \citep{anderson2004integrated}is a widely used approach to a fully unified cognitive architecture. The development of ACT-R is closely related to the latest research results of neurobiology. Compared to EPIC, the ACT-R (1) can only fire one production rule per cycle, (2) has a well-developed theory of declarative memory, and (3) contains learning mechanisms. Due to these advantages of ACT-R architecture, it is utilized in the experiment to examine the design.

\section{Proposed Design}

\subsection{Layout Design}

\begin{figure}
  \centering
  \caption{Layout of the proposed design: QWERT.}\label{fig.layout}
  \includegraphics[width=0.7\textwidth]{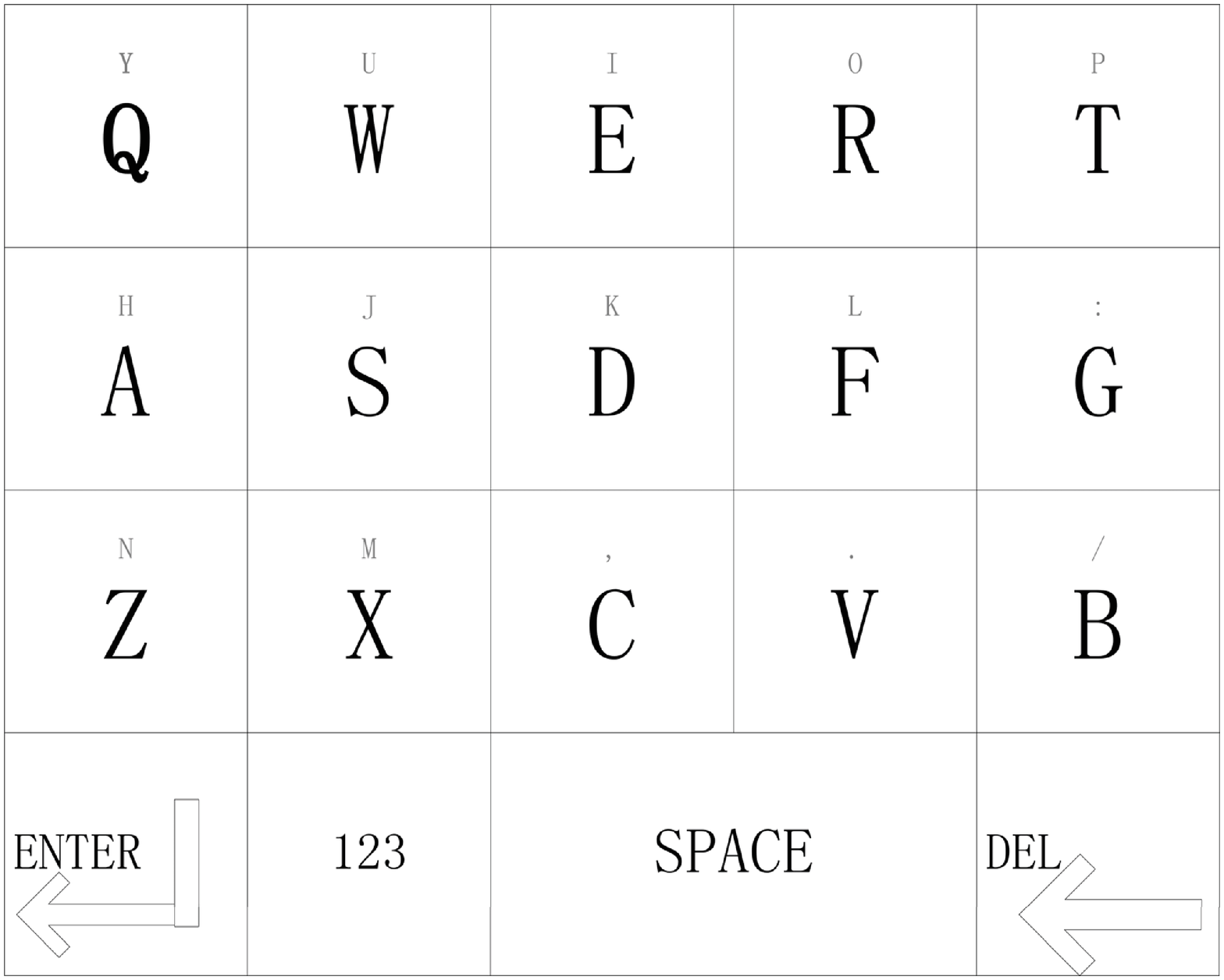}
\end{figure}

Considering the designs and recommendations given in Section 2, the layout of the proposed soft keyboard is shown in Figure \ref{fig.layout}. The details for the novel soft keyboard on touchscreen-based smartphones are presented as follows. \\

\textbf{(1)	Size}.
The size of each button and edge-to-edge spacing follows the recommendation given by \citet{park2010touch} and \citet{colle2004standing}, respectively. That is, each button size is around 10 mm and edge-to-edge spacing is 1 mm. The edge-to-edge spacing optional, in other words, 1 mm is used for large-sized touchscreens and waived for small-sized touchscreens. \\

\textbf{(2)	Layout}.
Instead of mirroring the right half of the keyboard to the left half proposed by \citet{matias1994half}, we duplicate the right half to the left. Because it has been shown that 57\% of the typing normally consists of characters from the left side of the normal keyboard, the left half is chosen as the main part. Take the first row as an example, we mainly use the left half of QWERT, i.e. Q, W, E, R, T, and take the right half as the auxiliary function, i.e. Y, U, I, O, P. The first row in the proposed soft keyboard encompasses the following series of keys Q/Y, W/U, E/I, R/O, T/P. Each button is assigned a double operation. Different operations trigger different letters as input. Extending this method to all relevant keys of the QWERTY, we derive a novel soft keyboard. Due to the first row of the proposed design, the novel soft keyboard is called QWERT (see Figure \ref{fig.layout}). \\

\textbf{(3)	Using Method}.
As proposed by \citet{matias1994half}, the choice of character depends on the time depression of a specific key or the force with which the key is depressed. However, this method is improper for our design which is based on touchscreen. Nevertheless, we can make use of the specific characteristics of touchscreen-based smartphones. In other words, the determining parameter can depend on users' gestures on the touchscreen. That is, the input of the left half letters of QWERTY (big font in Figure \ref{fig.layout}) is activated when users touch or tap the corresponding button (for example, "Q") while the transmission of the right half (small font letter is activated when users slide up the button ("Y" in this case). The last row contains several frequent operations, enter, number, space and delete. The enter, space and delete function are activated by simply tapping. The tapping on the number button will trigger a small $3\times3$ numeric keypad for users to input numbers. Since the analysis of the cognitive architecture of the number button is complicated, the discussion of the number button is omitted in this paper. \\

\begin{figure}[htbp]
  \centering
  \caption{Holding methods using left hand (A) and right hand (B).}\label{fig.holding}
  {A B}\\
  \includegraphics[width=0.4\textwidth]{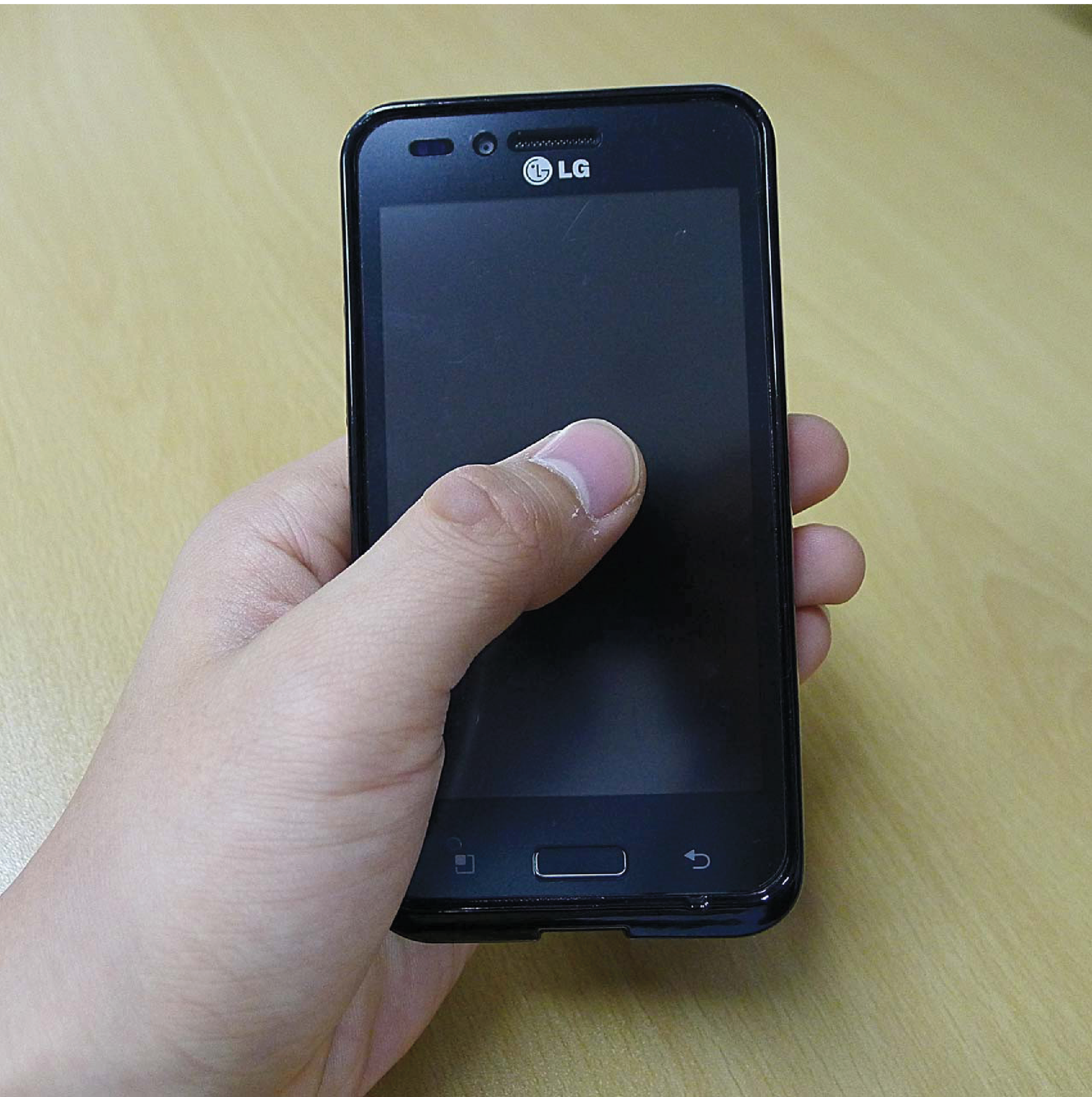}
  \includegraphics[width=0.4\textwidth]{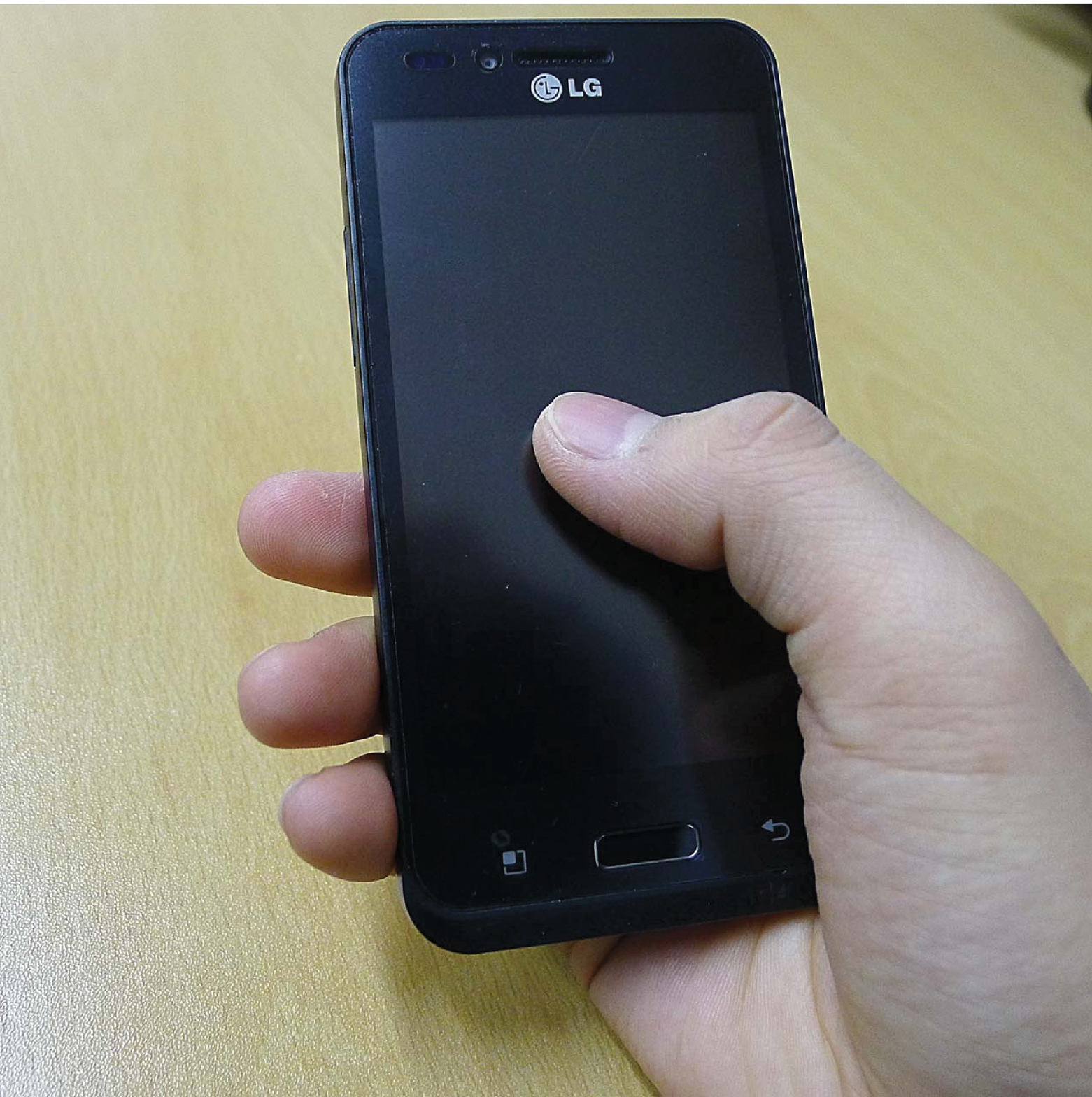}
\end{figure}

\textbf{(4) Phone holding method}.
In this paper, we assume the users are accustomed to use thumbs for text input, including left-handed and right-handed. Two common holding examples using LG LU6800 are shown in Figure \ref{fig.holding}. \\

\subsection{Application on Android OS}

\begin{figure}
  \centering
  \caption{Application of QWERT keyboard on Android.}\label{fig.andr}
  \includegraphics[width=0.5\textwidth]{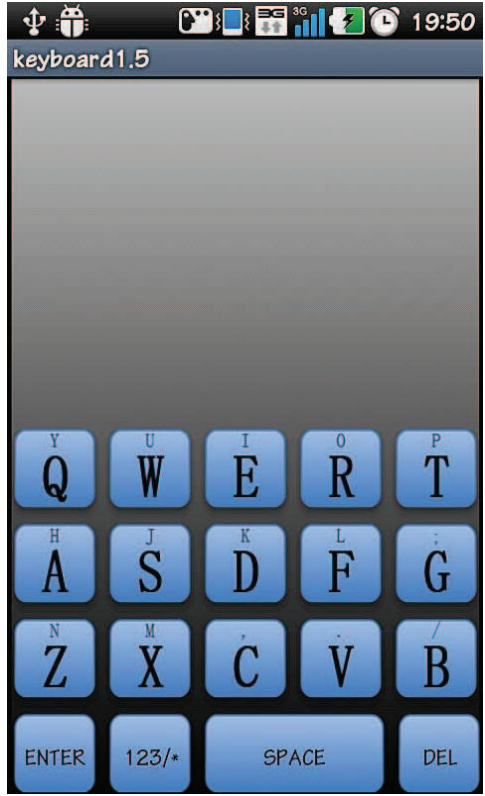}
\end{figure}

The proposed QWERT is implemented as a plugin on an Android 2.3.3 device, LG LU6800, as shown in Figure \ref{fig.andr}. For research convenience, we develop a text display area on the application. When a user input letters with the QWERT, the letter will display at the cursor. Text input requests, such as message input or website input, trigger the soft keyboard to appear. \\

\begin{figure}
  \centering
  \caption{Tapping and sliding methods of QWERT keyboard on Android.}\label{fig.show}
  \includegraphics[width=0.7\textwidth]{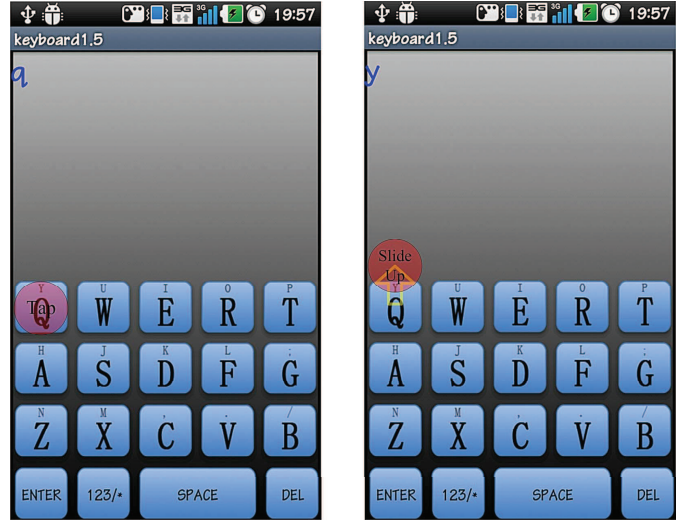}
\end{figure}

As we mentioned earlier, touching the button triggers the input of the corresponding letter while sliding up triggers the upper letter on the button. Figure \ref{fig.show} shows these two situations on the application. \\

\section{Empirical Evaluations}

Two methods are utilized for examining the performance of QWERT. The first experiment is examined on inexperienced subjects. Since there are no experienced users before releasing our application, a cognitive architecture model is used to simulate this situation in the second experiment. The purpose of these experiments is to examine the effectiveness of the proposed QWERT soft keyboard. \\

\subsection{Experiments on Inexperienced Users}

\subsubsection{Subjects}

Twenty male students from various majors at Korea University participated in the experiments. Their ages range from 21 to 28 years with the mean age of 24 and standard deviation of 1.8. All participants are right-handed and able to move their thumbs freely. They have experience in using touch phones with 1.4 years on average. We assume that the region located on touch phones is a square and that half thumb is used to operate and move on touch smartphones. The subjects' finger widths were measured in terms of half of the distal length. According to the statistics of our subjects' fingers, the mean width of participants' half thumbs is 16 mm with standard deviation of 0.9 mm, as listed in Table \ref{tab.sub}. \\

\begin{table}
\centering
\caption{Statistics of ages and width of half thumbs for 20 participants.}\label{tab.sub}
\begin{tabular}{ccc}
  \hline
    & Age & Thumb (mm) \\
  \hline
  Mean & 24.2 & 16 \\
  Standard Deviation & 1.79 & 0.92 \\
  \hline
\end{tabular}
\end{table}

\subsubsection{Experimental Equipment}

In this section, the experimental prototype is examined on an Android smartphone model called LG LU6800 which has a 4.3 inch (3.7$\times$2.13 inch) touchscreen. Two default keyboards, QWERTY and 3$\times$4 keyboard, in LU6800 are used to compare with the proposed design. Table \ref{fig.sub} summarises the detailed information of three keyboards. Since the size of buttons of 3$\times$4 keyboard is larger than those of QWERT, users, especially those who have large thumbs, can tap the buttons easier. But they have to multi-tap the button in order to input some letters. For example, the users have to tap four times on the button of 7 if they desire to input the letter "s". This operation is inconvenient compared to QWERTY keyboard. As shown in the third column, the discrepancies of proposed design are two-fold. First, the button size is 10.2$\times$10.7 mm which satisfies the recommendations given in recent studies. Second, both tapping and sliding are used to choose the input letters, which makes full use of the property of touchscreens. \\

\begin{sidewaystable}
\centering
\caption{Features of three distinct keyboards: $3\times4$, QWERTY and QWERT.}\label{fig.sub}
  \includegraphics[width=\textwidth]{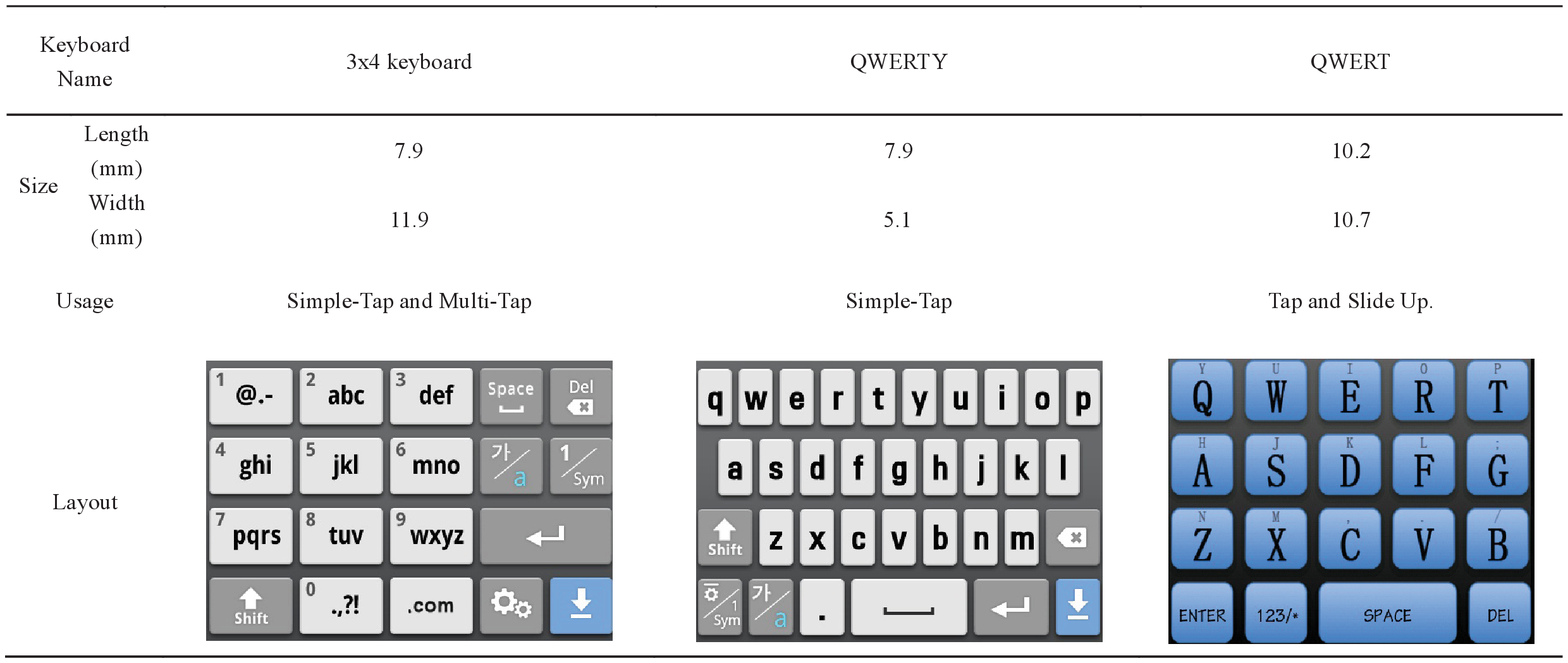}
\end{sidewaystable}

\subsubsection{Experiment Design}

Each human subject performed 10 sessions and only one session was allowed per day. 10 news articles from BBC were used as experimental text. These articles had been preprocessed so that they contained only lower case letters and simple punctuation, such as comma and period, to simplify the input. Each subject was forced to input the correct text as quickly as possible during each session using both right thumb and left thumb in ten minutes. Characters input per minute were recorded. Identical procedures were carried out on QWERT, QWERTY and 3$\times$4 keyboard. \\

\subsubsection{Results}

\begin{figure}
  \centering
  \caption{Input speed using QWERT, QWERTY and 3$\times$4 keyboard with left and right thumb over 10 sessions.}\label{fig.perf}
  \includegraphics[width=0.7\textwidth]{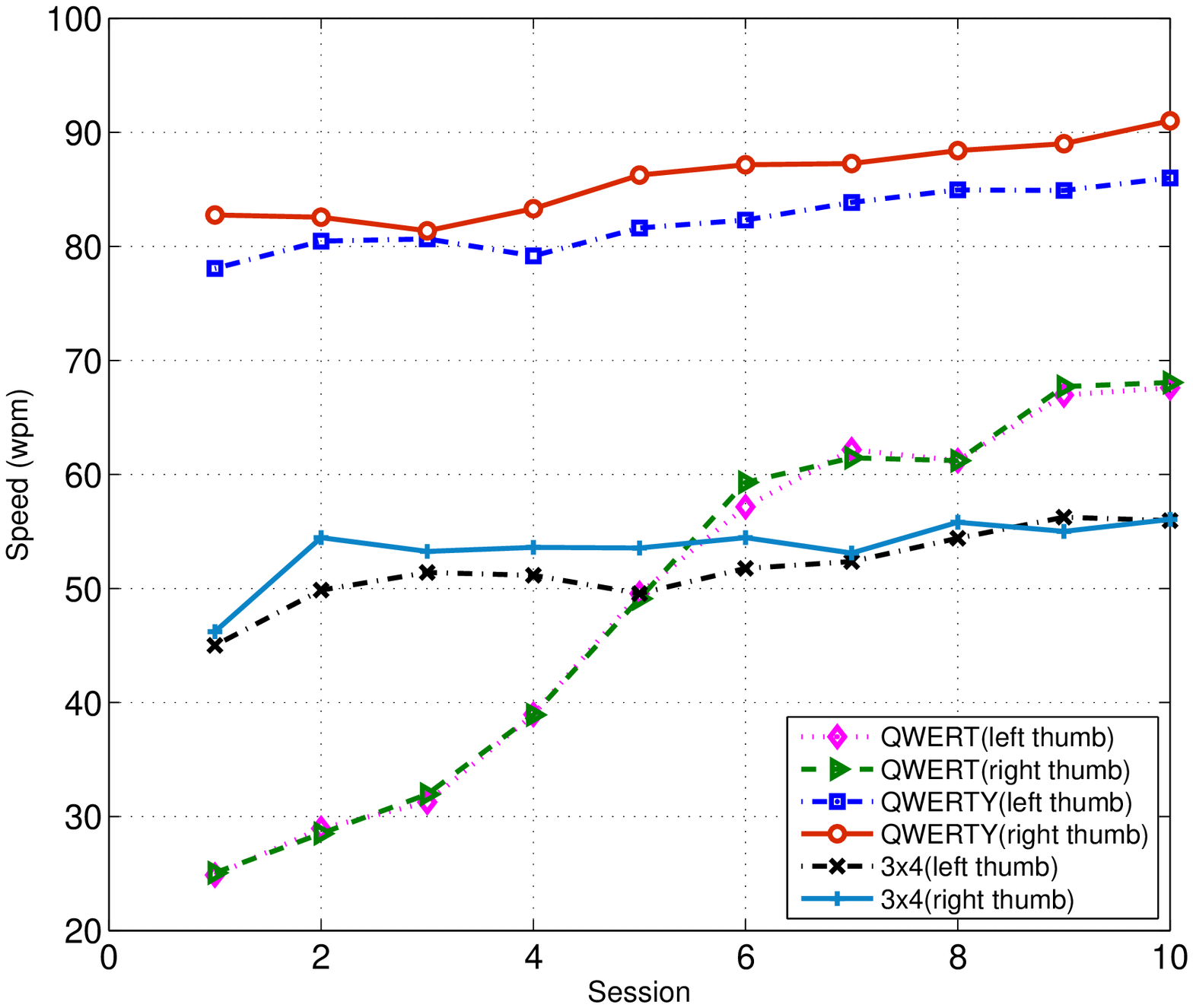}
\end{figure}

Figure \ref{fig.perf} plots the mean number of words input per minute (wpm) over 10 sessions by using three keyboards. The standard definition of a word as five characters is employed, thus wpm is obtained by multiplying characters per second by 60 and dividing by 5 \citep{mackenzie2002text}. From the depicted curve, we observe that the input speed of QWERT (using either left thumb or right thumb) improves obviously along 10 sessions and achieves 67.8 wpm at the 10\textit{th} session. It is noteworthy that the performance in the first session (where the average speed is 25 wpm) is impressive because no participants were trained before the experiment. This result is due to human's familiarity with original QWERTY. Another observation is that the performance using left thumb and right thumb is roughly identical in the curve. Thus we can conclude that the proposed design is efficient for both right-handed and left-handed users. Since the subjects are familiar with QWERTY, the input speed using QWERTY is the highest among three keyboards. In order to compare on the experienced users, we have to apply a cognitive architecture, ACT-R,  to stimulate the experiment. \\

\subsection{Experiments on Experienced Users}

\subsubsection{Methods}

\begin{figure}
  \centering
  \caption{Structure of ACT-R on QWERT.}\label{fig.str}
  \includegraphics[width=0.7\textwidth]{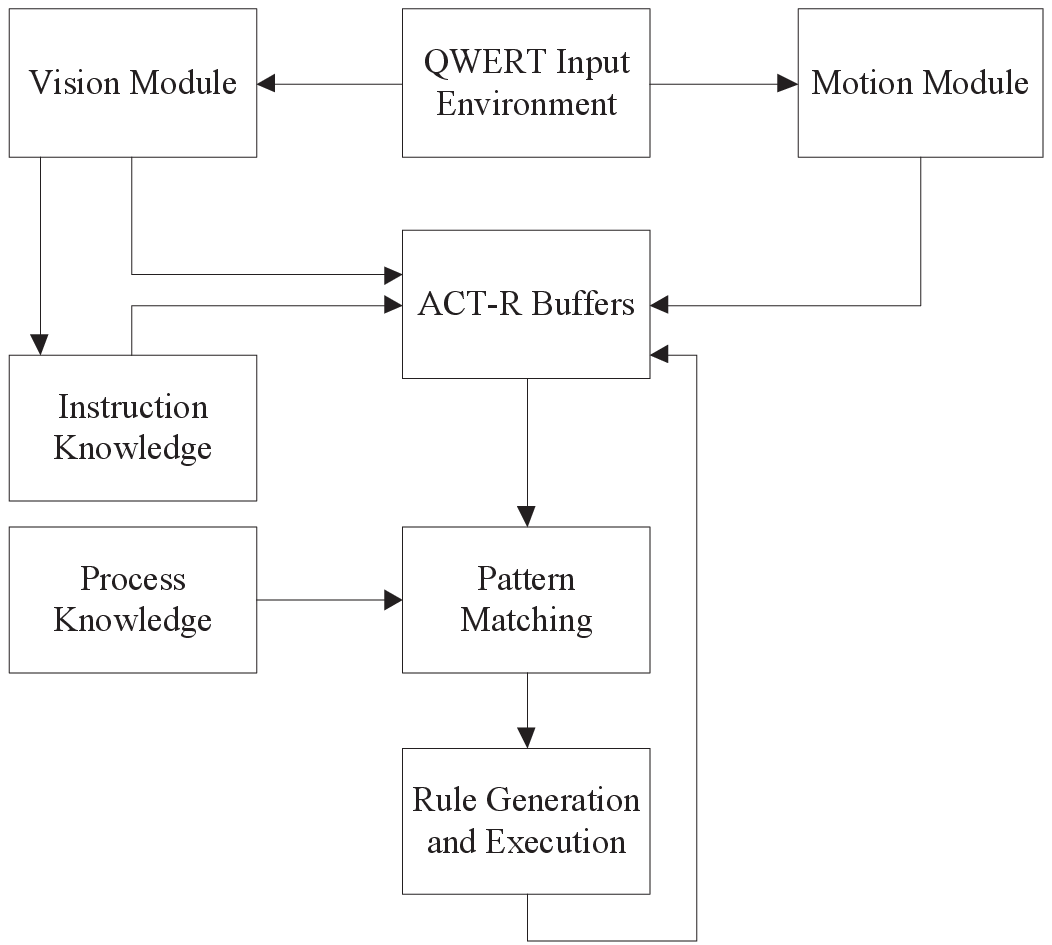}
\end{figure}

The movement time of experienced users who are familiar with the QWERT is different from inexperienced ones. There are several cognitive measures to predict the movement time. As addressed in Section 2, the ACT-R can be utilized directly in research domains instead of merely psychological area. The structure of ACT-R used in our experiment is depicted in Figure \ref{fig.str}. \\

In order to predict human movement time, the Fitts' Law is also built into ACT-R. But the parameters for Fitts' Law are estimated rather than measuring individuals. The reason is that the performance of a system is supposed to be predicted before the system is built. According to \citet{byrne2001act}, ACT-R uses the same Fitts' Law parameter estimates as the EPIC architecture does. In our experiment, the movement execution time is governed by a modified formulation of the Fitts' Law:
\begin{equation}
  MT=I_m \log_2(A/W+0.5)
\end{equation}
where $I_m=100[70\sim120]ms/bit$, $A$ is the distance between two buttons and $W$ is the width of the buttons. The motor module in ACT-R uses the Fitt's Law to calculate pointing times from the knowledge of where cursor was left by the last action, distance to and size of the next target in the keyboard. For example, if the target is clicked with a mouse, the predicted time is a combination of the Fitts' Law movement time plus the time to click the button. Though the Fitts' Law was originally established with a person tapping on a pad with a stylus, it is also available on a touchscreen. \\

\subsubsection{Experiment Design}

In order to build the cognitive architecture module, a general purpose UI prototyping tool based on ACT-R, CogTool, is used to construct our experiment. CogTool is a user interface (UI) prototyping tool that can produce quantitative predictions of how users behave when the prototype is implemented. We can rapidly analyze different products as part of competitive analysis and compare new designs with existing versions to ensure that the new design is better than others. CogTool's predictions are based on extensive prior research in cognitive psychology. Recent researches of users' tasks on mobile devices have been migrated into the released version of CogTool. CogTool is able to predict total execution time for an experienced user who performs a particular sequence of actions on a system. In our experiment, two kinds of human movements are defined: Moving and Sliding-up on the button. Moving is simulated by a default action on CogTool, i.e. "Move and Tap". Since there is no default setting for simulating sliding on CogTool, the "Down-tap" and "Up-tap" plus extra sliding time are used to simulate Sliding-up on the button. \\

In addition, "Think" steps, which are at the core of the research of CogTool, are inserted automatically because previous studies in the area of psychology and human-computer interaction have shown that humans need time to remember which button to press next when tapping on a button. The default time of the "Think" step in CogTool is 1.2s, which is too long for our case. Thus we set the "Think" step duration to 0.2s for 3$\times$4 keyboard, whereas 0.5s for QWERTY and QWERT. The reason is that more than one letter is placed on one button in 3$\times$4 keyboard, which is reasonable to reduce the "Think" time. For example, the user immediately double taps 3$\times$4 keyboard when he wants to input "b". Obviously, the "Think" step between the double taps on the same button needs less time. In either QWERTY or QWERT, the user has to move to a new key button whenever he wants to input a new letter. Thus the "Think" step duration is longer than 0.2s, as set 0.5s in our experiment. \\

In short, each moving and tapping on the button motion consists of eye movement preparation, eye movement, finger movement and the "Think" step. Following the construction, the text shown in Table \ref{tab.box} is tested using three distinct keyboards. \\
\begin{table}
\centering
\caption{Testing text for ACT-R.}\label{tab.box}
\begin{tabular}{|c|}
  \hline
  thanks for your dinner. take care. \\
  \hline
\end{tabular}
\end{table}

\subsubsection{Results}

Table \ref{tab.res} lists the predicted movement time in the simulated experiment on experienced users. The performance of QWERT outperforms both QWERTY and 3$\times$4 keyboard by comparing the spending time of inputting given text. The 3$\times$4 keyboard performs worst in this experiment since the multi-tapping is time consuming compared to other two methods. Unexpectedly, the proposed QWERT performs better than QWERTY. Two aspects may account for this result. (1) The design of the number of buttons and the size of each button is more convenient for users. (2) The usage of sliding on touchscreens which improves the interaction between users and smartphones. \\
\begin{table}
\centering
\caption{Spending times by using three keyboards.}\label{tab.res}
\begin{tabular}{cc}
  \hline
  Keyboard & Cost(s) \\
  \hline
  QWERTY & 16.628 \\
  3$\times$4 keyboard & 19.318 \\
  QWERT & 10.061 \\
  \hline
\end{tabular}
\end{table}

\section{Conclusions and Discussions}

	The popularity of touchscreen-based smartphones will continue to grow in the following decades. The users will still require for fast input methods on smartphones. To address this issue, a novel soft keyboard called QWERT is proposed for touchscreen-based smartphones in this paper. The QWERT is intuitively designed based on people's familiarity with traditional physical keyboards. We make use of the tactile characteristics of touchscreens, the gestures of tapping and sliding are input methods on QWERT. Furthermore, the size of each key button is rigorously arranged according to recent research results. In order to examine the effectiveness, a human subject experiment and a simulation experiment are conducted for checking different performance on experienced and inexperienced users, respectively. Both results show that using the proposed design results in faster input speed than the default QWERTY and 3$\times$4 keyboard in the touch smartphones. The QWERT is a viable option to improve user experience for smartphone developers. \\

A shortcoming of current QWERT is the neglection of influence of numbers, punctuations and letter cases. The design considering those characters and the experiment evaluating those characters are subject to a future study. With regards to the experiments, much more testing of the new keyboard on larger and different groups of human subjects should be taken into account in the future research.

\section*{Acknowledgement}
The author gratefully acknowledges the China Scholarship Council (CSC) for fellowship support.


\bibliographystyle{plainnat}
\bibliography{qwert}

\end{document}